\documentclass[aps, pra, twocolumn, tightenline, showpacs]{revtex4}

\usepackage{graphicx} 
\usepackage{epstopdf} 
\usepackage{epsfig}
\begin{document}

\title{Tunable Up-Conversion Photon Detector}
\author{R.~T.~Thew}\email{Robert.Thew@physics.unige.ch}

\author{H.~Zbinden}

\author{N.~Gisin}

\affiliation{Group of Applied Physics, University of Geneva, 1211 Geneva 4, Switzerland}

\date{\today}

\begin{abstract}
We introduce a simple approach for a tunable up-conversion detector. This scheme is relevant for both single photon detection or anywhere where low light levels at telecom wavelengths need to be detected with a high degree of temporal resolution or where high count rates are desired. A system combining a periodically poled Lithium niobate waveguide for the nonlinear wavelength conversion and a low jitter Silicon avalanche photodiode are used in conjunction with a tunable pump source. We report more than a ten-fold increase in the detectable bandwidth using this tuning scheme.
\end{abstract}


\maketitle

\newpage

Single photon detection, or indeed, detection of any low light signals at telecommunication wavelengths, has suffered from a variety of performance constraints. The reliance on InGaAs/InP avalanche photodiodes (APDs) has meant, until recently, working with efficiencies of around 10\% in a gated, or triggered, regime with relatively high levels of noise \cite{Stucki01} compared to Silicon (Si) APDs \cite{Ghioni03}. Despite this, many seminal experiments in quantum cryptography, see \cite{Gisin02a}, have helped push this technology to the level of a commercial viability \cite{ComInGaAs}.  The introduction of single photon counting for distributed telecommunication measurements, such as optical time domain reflectometry (OTDR) has also provided significant advantages in terms of sensitivity and precision \cite{Scholder02a}. Furthermore, whether it be in telecommunication, where faster and faster communication results in a lower mean number of photons per bit (pulse), or more generally in any low light-level metrology scheme, the role of single photon detection is increasing in importance. 

Recently, we have seen the arrival of detectors based on superconduction \cite{Korneev04} that hold great promise. These detectors have low efficiencies, for the moment, of less than 10\% at 1550\,nm, although this is off-set by their potential for: very low noise, a few Hz; low timing jitter (temporal response) of less than 20ps; and with improvements in electronics, count rates approaching 1\,GHz \cite{Korneev04}. Their drawback is, however, the need for cryostatic cooling. An alternative approach that has been pursued by a few groups \cite{VanDevender04a, Roussev04, Albota04a, Thew06}  is to combine nonlinear up-conversion, also referred to as sum-frequency generation (SFG), to convert the telecom wavelength photons to the visible spectrum where Si APDs can be harnessed.  It should be mentioned that this is not restricted to this regime and some of us have previously shown its operation in the mid-IR at 4.6\,$\mu$m \cite{Karstad05}. These approaches also have the added advantage of bringing a passive detection technique to these regimes. In our particular case we normally use either one of two different types of Si APDs (MPD: PD5CTA, id Quantique: id100-50)  that provide for very low timing jitter ($<$\,50\,ps) \cite{Cova89a, Rochas03a}. This is an order of magnitude improvement over both InGaAs/InP \cite{Stucki01} and the standard Si APDs \cite{Ghioni03} currently in wide use. We have already used these up-conversion detectors to significantly increase transmission rates for QKD \cite{Thew06} and, more recently, gain another order of magnitude increase in the 2-point, fault-finding, resolution  for a single-photon OTDR scheme \cite{Legre07a}. 
\begin{figure}[!h]
\begin{center}
\epsfig{figure=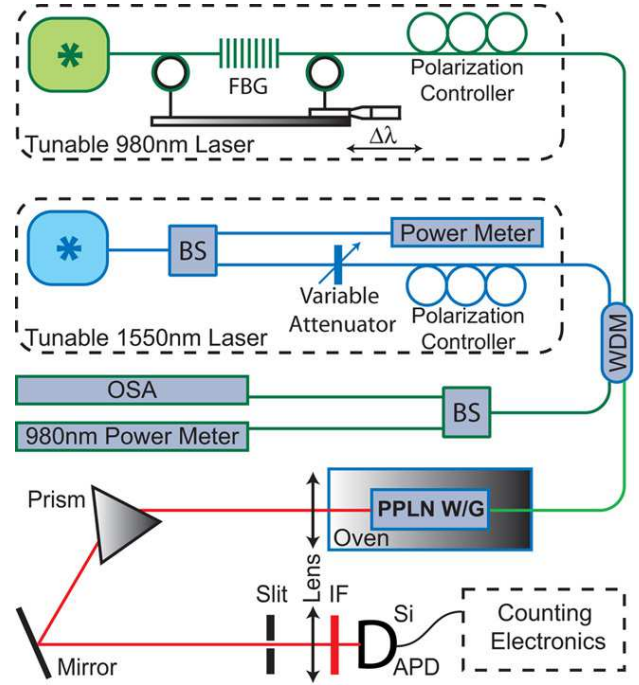,width=8.2cm}
\caption{Experimental scheme: (Single) Photons at 1550\,nm are combined with a tunable (see text) laser at 980\,nm  in a fibre WDM before entering a nonlinear PPLN waveguide for up-conversion to 600\,nm. This output is filtered using a prism and an interference filter (IF) before detection with a Si APD. Signal and pump, intensities and wavelengths (OSA - optical spectrum analyser), after fibre beamsplitters (BS).}
\label{fig:schema}
\end{center}
\end{figure}

Due to the narrow acceptance bandwidth, governed by the nonlinear conversion process, these up-conversion scheme has been restricted to operating in systems with well defined source wavelengths such as in QKD and OTDR. In this Letter we introduce a variation to the up-conversion detection scheme that gives over a ten-fold increase for the detectable bandwidth. We firstly introduce the basic principle and operation in the context of recent improvements in device performance before detailing the process used for tuning the detection wavelength. We finish by elaborating on possible extensions to this idea.

The detection scheme we propose is illustrated in Fig. \ref{fig:schema}. The underlying principle involves the nonlinear up-conversion of a signal in the telecom band, around 1550\,nm, to the visible regime, followed by its subsequent detection with a Silicon (Si) APD. In more detail, the signal photons are mixed at a wavelength division multiplexor (WDM) with a strong {\it pump} laser at 980\,nm. We will describe the tuning mechanism in more detail momentarily. This fibre is pigtailed to a temperature stabilised periodically poled Lithium niobate (PPLN) waveguide (W/G) (HC Photonics) where the nonlinear wavelength conversion is performed. The waveguide is 2.2\,cm long, and has a poling period of 9\,$\mu$m with a normalised internal efficiency of over 500\%\,W$^{-1}$\,cm$^{-2}$. After the waveguide, the up-converted light is collimated and passed through a filtering system, consisting of a prism and an interference filter, centered at 600\,nm to remove any excess pump photons and their second harmonic generation (SHG) signal that is also present. Finally, we focus the signal onto a free space Si APD (MPD: PD5CTA). Note, that this nonlinear process is equally valid for classical-level light pulses and even down to the level of a single photon.

\begin{figure}[!t]
\begin{center}
\epsfig{figure=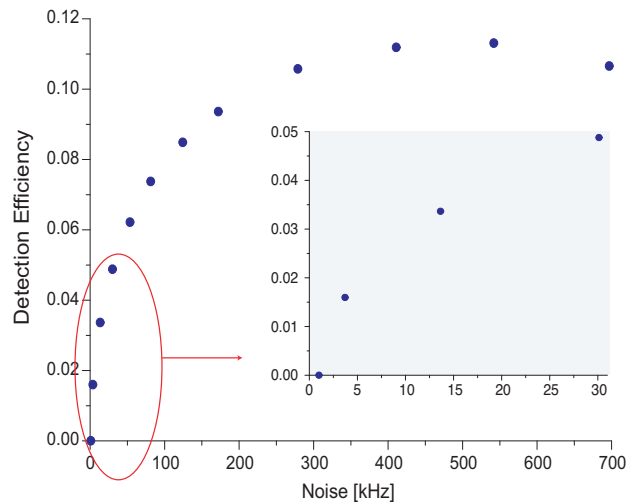,width=8.2cm}
\caption{The efficiency-noise characteristics for the up-conversion detector showing a peak efficiency of nearly 12\,\%. Inset we see the low noise efficiency response in detail.}
\label{fig:efficiency}
\end{center}
\end{figure}

Since our first attempts at this type of detection \cite{Thew06}, improvements in the fabrication of the PPLN W/Gs, the filtering, as well as optimization of the Si APDs for these schemes has seen overall detection efficiencies greater than 10\% obtained.  The efficiency-noise characteristics are a function of the pump power and are shown in Fig. \ref{fig:efficiency}. This is the efficiency for obtaining an electrical output, a click, when we send in a 1550\,nm photon.  We see clearly that in this instance that significant noise persists, as was the case for all previous experiments \cite{VanDevender04a, Albota04a, Thew06}. Inset, we see a close-up of the more usable, lower efficiency, characteristics - depending on the requirements one can sacrifice efficiency for a lower noise level as is commonly done.

The signal wavelength detected is determined by the  pump wavelength and the quasi-phase matching (QPM) condition of the nonlinear crystal. This condition is described by \cite{deMicheli97}
\begin{eqnarray}
\frac{n(s)}{\lambda(s)} + \frac{n(p)}{\lambda(p)} + \frac{m}{\Lambda} = \frac{n(u-c)}{\lambda(u-c)}
\end{eqnarray}
where $n$ is the effective refractive index in the waveguide, $\lambda$ denotes the different wavelengths, $s$ and $p$ refer to the signal and pump respectively and $u-c$ stands for up-conversion. Finally, $\Lambda$ is the poling period for the $m$-th order quasi-phase matched condition of the nonlinear PPLN W/G. As previously mentioned, the nonlinear interaction imposes a constraint on the detection bandwidth and hence all previous systems have worked only for very well defined wavelengths. We wish to increase the range of wavelengths that can be used by such a detector to improve the practicality of these devices. As we can see there are several possibilities to {\it tune} the detector to a desired detection wavelength. One can change the temperature, thus using the temperature dependence of the refractive indicies for the different wavelengths. Different poling periods for the phase matching could also be incorporated or, as we have chosen to do here, one can change the pump wavelength. 

While the first two choices are feasible they are not particularly practical. Changing the temperature to tune the QPM is possible, where a 10\,K shift in temperature changes the accepted QPM signal wavelength by around 3-4\,nm. Unfortunately this is a very slow process where the speed and  stability for changing wavelengths is governed by the rate at which thermal equilibrium can be recovered. The poling period is something that needs to be determined at the production stage, but samples are commonly fabricated with series of differently poled regions. Hence, one could imagine moving the different W/Gs in and out of the optical beam to choose the desired interaction. This is however, both very slow, and given the difficulty in alignment of these devices, highly impractical. 

Our original choice of components was made with a view to simplicity and the pump that we use is a standard fibre coupled 980\,nm laser diode as used, for example, in telecom Erbium doped amplifiers. The lasing wavelength of these devices is determined by the reflection band from an external cavity consisting of a fibre Bragg grating. Thus,  varying the central wavelength reflected by the Bragg grating tunes the wavelength of our pump and hence that of the detected signal. We use a fibre stretcher to physically lengthen the fibre Bragg grating and hence change the grating periodicity and subsequently, the pump laser wavelength. 

\begin{figure}[!t]
\begin{center}
\epsfig{figure=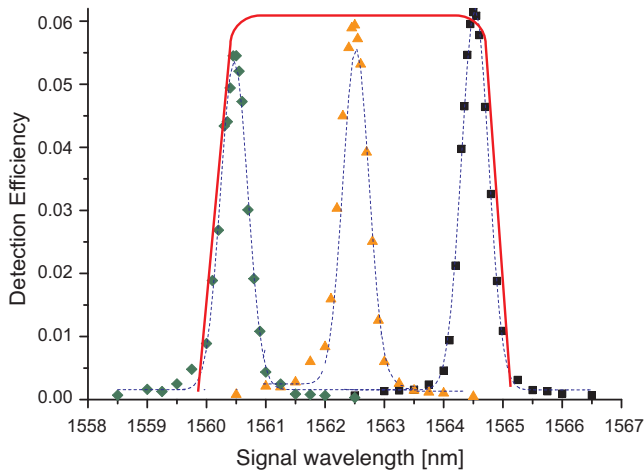,width=8.5cm}
\caption{The detection efficiency as a function of the signal wavelength for three different pump wavelengths. The peak efficiency corresponds to a noise level of 50\,kHz in each case. We see the normal acceptance bandwidth of $<$0.5\,nm can be extended ten fold to around 5\,nm as illustrated by the envelope - this is to guide the eye and is not a fit.}
\label{fig:multipeaks}
\end{center}
\end{figure}

We see in Fig. \ref{fig:multipeaks} the efficiency as a function of the signal wavelength for three different pump wavelengths. In principle we can tune continuously over this range but have shown discreet values to give an indication of the fundamental acceptance bandwidth of the QPM. One must ensure that the bandwidth of the WDM is also sufficient to separate the scanned pump and signal wavelengths. This measurement is realised by first selecting the pump wavelength, which we monitor after the WDM (see Fig.  \ref{fig:schema}) with an optical spectrum analyser (OSA) and power meter, and then choosing a pump power that corresponds to around 6\,\% detection efficiency and 50\,kHz noise. The signal photons are generated by a CW tunable laser (Exfo: FLS-2600B) and attenuator (Exfo: FVA-60B) and the wavelength is scanned to find the optimal conversion efficiency. This efficiency is calculated simply by the number of photons detected at 600\,nm with respect to the number at 1550\,nm after the variable attenuator shown in Fig. \ref{fig:schema}.

We also see in Fig. \ref{fig:multipeaks} the acceptance bandwidth for the signal is around 0.5\,nm. We have drawn an envelope over the three curves to illustrate the gain in useable detection bandwidth that this scheme provides. In this instance we were quite conservative about tuning the pump wavelength, 2\,nm. Realistic limits of around 4-5\,nm would result in a overall acceptance bandwidth of $>$\,10\,nm. If one takes this idea a little further, we could imagine an integrated array of such detectors. This could be realised with a WDM on the input fibre, separating the incoming signal into smaller bands, pigtailed to the same PPLN sample, with multiple waveguide zones. Each band would require their own pump and certainly the cost and complexity of the overall system would increase, but 4 pumps could see the detection bandwidth cover the whole telecom C-band


In conclusion, we have presented a simple scheme for a compact and tunable single-photon telecommunication wavelength detector capable of passive operation and with high count rates and timing resolution.

\section*{Acknowledgements}
RTT would like to thank Alexios Bevaratos for useful discussions concerning the original idea and the authors  acknowledge financial support from the European project SECOQC and QAP and the
Swiss NCCR "Quantum Photonics".


\begin{references}


\bibitem{Stucki01} D.~Stucki, G.~Ribordy, A.~Stefanov, H.~Zbinden, J.~G.~Rarity, T.~Wall, J. Mod. Opt., {\bf 48}, 1967 (2001)

 \bibitem{Ghioni03} M.~Ghioni, A.~Guidice, S.~Cova and F.~Zappa, J.Mod. Opt., {\bf 50} 2251 (2003)
 
\bibitem{Gisin02a} N.~Gisin, G.~Ribordy, W.~Tittel and H.~Zbinden, Rev. Mod. Phys., {\bf 74} 145 (2002)
 
\bibitem{ComInGaAs} www.idquantique.com

\bibitem{Scholder02a} F.~Scholder, J-D.~Gautier, M.~WegmŸller and N.~Gisin, Opt. Comm., {\bf 213} 57 (2002)


\bibitem{Korneev04} A.~Korneev, P.~Kouminov, V.~Matvienko, G.~Chulkova, K.~Smirnov, B.~Voronov, G.~N.~Gol'tsman, M.~Currie, W.~Lo, K.~Wilsher, J.~Zhang, W.~Slysz, A.~Pearlman, A.~Verevkin, and R.~Sobolewski, App. Phys. Lett., {\bf 84} 5338 (2004)

\bibitem{VanDevender04a} A.~P.~VanDevender and P.~G.~Kwiat,  J. Mod. Opt., {\bf 51}, 1433 (2004)

 
\bibitem{Albota04a} M.~A.~Albota and F.~N.~C.~Wong, Opt. Lett., {\bf 29} 1449 (2004)

\bibitem{Roussev04} R.~V.~Roussev, C.~Langrock, J.~R.~Kurz and M.~M.~Fejer, Opt. Lett., {\bf 29}, 1518 (2004)

\bibitem{Thew06} R.~T.~Thew, S.~Tanzilli, L.~Krainer, S.~C.~Zeller, A.~Rochas, I.~Rech, S.~Cova, H.~Zbinden and N.~Gisin, New J. Phys., {\bf 8}, 32 (2006)

\bibitem{Karstad05} K.~Karstad, A.~Stefanov, M.~Wegmuller, H.~Zbinden, N.~Gisin, T.~Aellen, M.~Beck, J~Faist, Opt. \& Las. Eng., {\bf 43} 537 (2005)

\bibitem{deMicheli97} M.~P.~De~Micheli, Qu. SemiCl. Opt., {\bf 9} 155 (1997).

\bibitem{Cova89a} S.~Cova,  A.~Lacaita, M.~Ghioni, G.~Ripamonti and  T.~A~ Louis, Rev. Sci. Instr., {\bf 60}, 1104 (1989)

\bibitem{Rochas03a} A.~Rochas, M.~Gani, B.~Furrer, P.~A.~Besse, R.~S.~Popovic, G.~Ribordy and N.~Gisin Rev. Sci. Instr., {\bf 74} 3263 (2003)

\bibitem{Legre07a} M.~Legre, R.~Thew, H.~Zbinden, N.~Gisin, Opt. Exp., {\bf 15} 8237 (2007)



\end{references}
\end{document}